\documentclass{aa}  

\usepackage{graphicx}

\usepackage{txfonts}
\begin{document}

\title{Galactic Trajectories of Interstellar Objects 1\textit{I}/'Oumuamua, 2\textit{I}/Borisov,  and 3\textit{I}/Atlas}

   \subtitle{}

   \author{Shokhruz Kakharov
          \and
          Abraham Loeb
          }

    \institute{Astronomy Department, Harvard University, 60 Garden St., Cambridge, MA 02138, USA,\\
              \email{shokhruzbekkakharov@college.harvard.edu; aloeb@cfa.harvard.edu}
             }

   \date{}

 
 \abstract
{The first interstellar objects, 1I/`Oumuamua, 2I/Borisov and 3I/ATLAS, were discovered over the past decade.}
{We follow the trajectories of known interstellar objects in the gravitational potential of the Milky Way galaxy to constrain their possible origin.}
{We perform Monte Carlo orbital integrations using 10,000 trajectory ensembles per object to properly account for measurement uncertainties in both object velocities and Solar motion parameters. We implement a Bayesian statistical framework that combines a Rayleigh-like likelihood function with star formation rate priors to infer stellar ages from the maximum vertical excursions ($z_{\text{max}}$) of orbital trajectories. The likelihood function incorporates age-dependent velocity dispersions reflecting the thin-thick disk transition and dynamical heating over galactic history.}
{Our Monte Carlo analysis yields median $z_{\text{max}}$ values of 0.016 $\pm$ 0.002 kpc for 1I/`Oumuamua, 0.121 $\pm$ 0.010 kpc for 2I/Borisov, and 0.480 $\pm$ 0.020 kpc for 3I/ATLAS. The Bayesian age inference indicates that 1I/`Oumuamua originated from a young stellar system (1.0 Gyr, 68\% CI: 0.1-4.1 Gyr), 2I/Borisov from an intermediate-age population (3.8 Gyr, 68\% CI: 1.8-5.9 Gyr), and 3I/ATLAS from an old thick-disk source (9.6 Gyr, 68\% CI: 7.8-10.3 Gyr). These results demonstrate clear age discrimination where smaller vertical excursions correspond to younger stellar origins.

}
   {}

   \keywords{interstellar objects --
                galactic trajectories --
                origins
               }

   \maketitle
%

\section{Introduction}

   The discovery of interstellar objects over the past decade has sparked significant interest in understanding their origins and dynamics (see reviews by \citet{2022AsBio..22.1459S}, \citet{2023ARA&A..61..197J}, \citet{2024arXiv240706475J}, and references therein). A fundamental unknown is the likely origin of each of these objects~\citep{2018AJ....156..205B,bailer2020}. Constraining the  sources could shed light on the nature of these interstellar objects and the astrophysical processes that created them (see, for example,~\citet{2020NatAs...4..852Z} or~\cite{2024A&A...686A.123L}).

In this paper, we numerically integrate the trajectories of interstellar objects back in time in the gravitational potential of the Milky-Way as a way to relate them to potential stellar populations. For simplicity, we ignore transient gravitational features such as spiral arms and the Galactic bar. This is a reasonable approximation for orbits in the outer part of the Galactic disk. 

We implement a comprehensive Monte Carlo approach that addresses observational uncertainties through ensemble orbital calculations. Rather than computing single deterministic trajectories using published velocity measurements at face value, we generate 10,000 orbital realizations per object by sampling from the measurement uncertainty distributions. This methodology properly propagates both the formal errors in interstellar object velocities (derived from JPL Small-Body Database covariance matrices) and systematic uncertainties in Solar motion parameters through the orbital calculations. Each Monte Carlo sample creates slightly different initial conditions, leading to a statistical distribution of maximum vertical heights ($z_{\text{max}}$) that quantifies the true uncertainty in our orbital predictions over gigayear timescales.

By integrating the orbits of these objects back in time, we are able to constrain the spatial region of their potential sources within the Milky Way. These constraints limit the possible birthplaces of the different interstellar objects and provide insights into the Galactic environment from which they originated.

Since the scale-height of stars in the Milky-Way disk increases with age, we use the vertical excursion of each interstellar object from the Milky-Way disk mid-plane to constrain its likely age. We develop a rigorous Bayesian statistical framework that combines the orbital $z_{\text{max}}$ measurements with theoretical models of stellar kinematics to infer the ages of the parent stellar systems. Our likelihood function adopts a Rayleigh-like distribution that emerges naturally from assuming stellar velocities follow three-dimensional Gaussian distributions combined with harmonic oscillator dynamics in the galactic potential. The age-dependent velocity dispersion model incorporates the observed thin-thick disk transition, with young stars exhibiting small dispersions due to limited dynamical heating and old populations showing large dispersions from billions of years of gravitational scattering. Any dynamical pumping of the stellar scale-height by gravitational perturbations from satellite galaxies or star clusters, would affect interstellar objects and stars alike since both populations are collisionless. Hence, our constraints apply to the full age of the interstellar objects irrespective of their travel time.

The organization of this paper is as follows. Section 2 describes our Monte Carlo orbital integration methodology and the resulting ensemble statistics for the Galactic trajectories of individual interstellar objects, with subsections dedicated to 2I/Borisov (\S 2.2), 1I/`Oumuamua (\S 2.3) and 3I/ATLAS (\S 2.4). In Section 3, we present our Bayesian statistical framework for age inference, including the theoretical foundation of the likelihood function, the age-dependent velocity dispersion model, and the resulting posterior age distributions for each object. Finally, we summarize the implications of our results in Section 4.

\section{Galactic Trajectories of Interstellar Objects}

\subsection{Method of Calculation}

Our numerical integration is based on the OrbitIntegrator from GalPot, which utilizes the MWPotential2014 model from McMillan (2017) to simulate the gravitational potential of the Milky Way galaxy. We assume a circular velocity of 233 km~s$^{-1}$ and an orbital radius of 8.3 kpc for the Local Standard of Rest (LSR) around the Galactic center, consistently with the latest Gaia data~\citep{2023A&A...676A.134P}. We initiate the trajectories from the velocity of each interstellar object relative to the LSR and follow them back in time for 10 Gyr, roughly the age of the Milky Way disk~\citep{2019ApJ...887..148F}. The code uses the NumPy's \texttt{linspace} function that defines the relevant timescale, and then integrates the orbits using the OrbitIntegrator's \texttt{getOrbitPath and Stats} method. The resulting orbit paths are extracted and plotted using Matplotlib, with separate plots showing the radial distance $R$, vertical excursion from the disk mid-plane $z$, and azimuthal angle $\phi$, as functions of time. The code also calculates the evolution of the velocity components $v_R$, $v_z$, and $v_\phi$ as functions of time. 

In order to convert the velocity measurements in the Solar system to the Galactic frame of reference, we incorporate the motion of the Sun relative to the Local Standard of Rest (LSR), with the Galactic components: $U_\odot = 10.79 \pm 0.56$ km s$^{-1}$, $V_\odot = 11.06 \pm 0.94$ km s$^{-1}$, and $W_\odot = 7.66 \pm 0.43$ km s$^{-1}$~\citep{2022A&A...667A..98R}. Before entering the solar system, 2I/Borisov's velocity was $(U, V, W) = (33.1, -6.8, 8.3)$ km s$^{-1}$, while 1I/`Oumuamua's velocity relative to the Sun is $(U-U_\odot, V-V_\odot, W-W_\odot) = (-11.457 \pm 0.009, -22.395 \pm 0.009, -7.74 \pm 0.011)$ km s$^{-1}$~\citep{bailer2020,2017RNAAS...1...21M}. For the interstellar object 3I/ATLAS (C/2025 N1), we adopt its inbound heliocentric velocities of $(U, V, W) = (-54.4, -20.3, +19.5)$ km s$^{-1}$ and corresponding Galactocentric cylindrical velocities of $(v_R, v_z, v_\phi) = (+43.62, +27.18, +223.76)$ km s$^{-1}$, derived from JPL orbital solution JPL\#5. We verified the orbital calculation accuracy through direct API comparison with JPL Horizons (achieving $<10^{-10}$ AU precision), independent mathematical validation of the elements-to-Cartesian conversion, and consistency checks with predicted planetary encounter geometries. The Galactocentric velocities were obtained via standard coordinate transformations incorporating solar motion parameters.

\subsubsection{Monte Carlo Uncertainty Propagation}

The fundamental limitation of single-orbit calculations is that they ignore observational uncertainties in both the interstellar object velocities and Solar motion parameters. To address this limitation, we implement a comprehensive Monte Carlo approach that properly propagates measurement errors through the orbital integrations.

Our Monte Carlo methodology involves generating 10,000 orbital realizations per object by sampling from the appropriate uncertainty distributions. For each sample, we perturb the nominal heliocentric velocities by adding random errors drawn from Gaussian distributions with standard deviations derived from the formal covariance matrices. We queried the JPL Small-Body Database for orbital element uncertainties and propagated these to velocity uncertainties in the Galactic frame. For 3I/ATLAS, this yields $\sigma_U = 0.011$ km s$^{-1}$, $\sigma_V = 0.012$ km s$^{-1}$, $\sigma_W = 0.020$ km s$^{-1}$. Similarly, for 1I/`Oumuamua we obtain $\sigma_U = 0.009$ km s$^{-1}$, $\sigma_V = 0.009$ km s$^{-1}$, $\sigma_W = 0.011$ km s$^{-1}$, and for 2I/Borisov $\sigma_U = 0.030$ km s$^{-1}$, $\sigma_V = 0.030$ km s$^{-1}$, $\sigma_W = 0.030$ km s$^{-1}$.

Additionally, we account for systematic uncertainties in Solar motion parameters by sampling from distributions with $\sigma_U = 1.2$ km s$^{-1}$, $\sigma_V = 2.1$ km s$^{-1}$, $\sigma_W = 0.6$ km s$^{-1}$ for the velocity components and $\sigma_{V_c} = 15.0$ km s$^{-1}$ for the galactic rotation speed. These systematic errors affect the transformation from heliocentric to galactocentric reference frames and represent uncertainties in our knowledge of Solar motion relative to the Local Standard of Rest.

For each Monte Carlo sample, we transform the perturbed heliocentric velocities to the galactic rest frame, integrate the resulting orbit for 1 Gyr using the identical potential model and numerical methods, and record the maximum absolute vertical excursion $z_{\text{max}}$ achieved during the integration period. This process yields a statistical distribution of $z_{\text{max}}$ values for each object that properly reflects the observational uncertainties.

The Monte Carlo ensemble results are summarized using robust statistical measures: the median $z_{\text{max}}$ as our best estimate and the 16th and 84th percentiles to define 68\% confidence intervals. This approach provides a rigorous framework for uncertainty quantification that is essential for subsequent Bayesian age inference.

\subsection{1I/'Oumuamua}

\begin{figure}[!ht]
    \centering
    \includegraphics[width=0.45\textwidth]{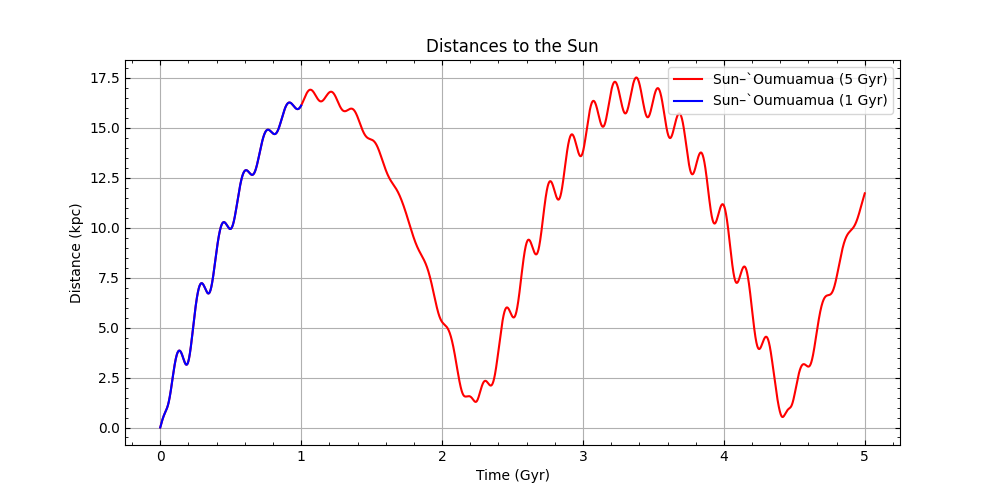}
    \caption{Distance of 1I/'Oumuamua from the Sun back in time.}
    \label{Fig:O1}
\end{figure}

The past evolution of the distance of the interstellar object 1I/`Oumuamua from the Sun is shown in figure~\ref{Fig:O1}. The 1I/`Oumuamua-Sun separation follows a period of about 2.2 Gyr, with 1I/`Oumuamua being on the other side of the Milky Way disk relative to the Sun about 1.1 Gyr ago.

\begin{figure}[!ht]
    \centering
    \includegraphics[width=0.5\textwidth]{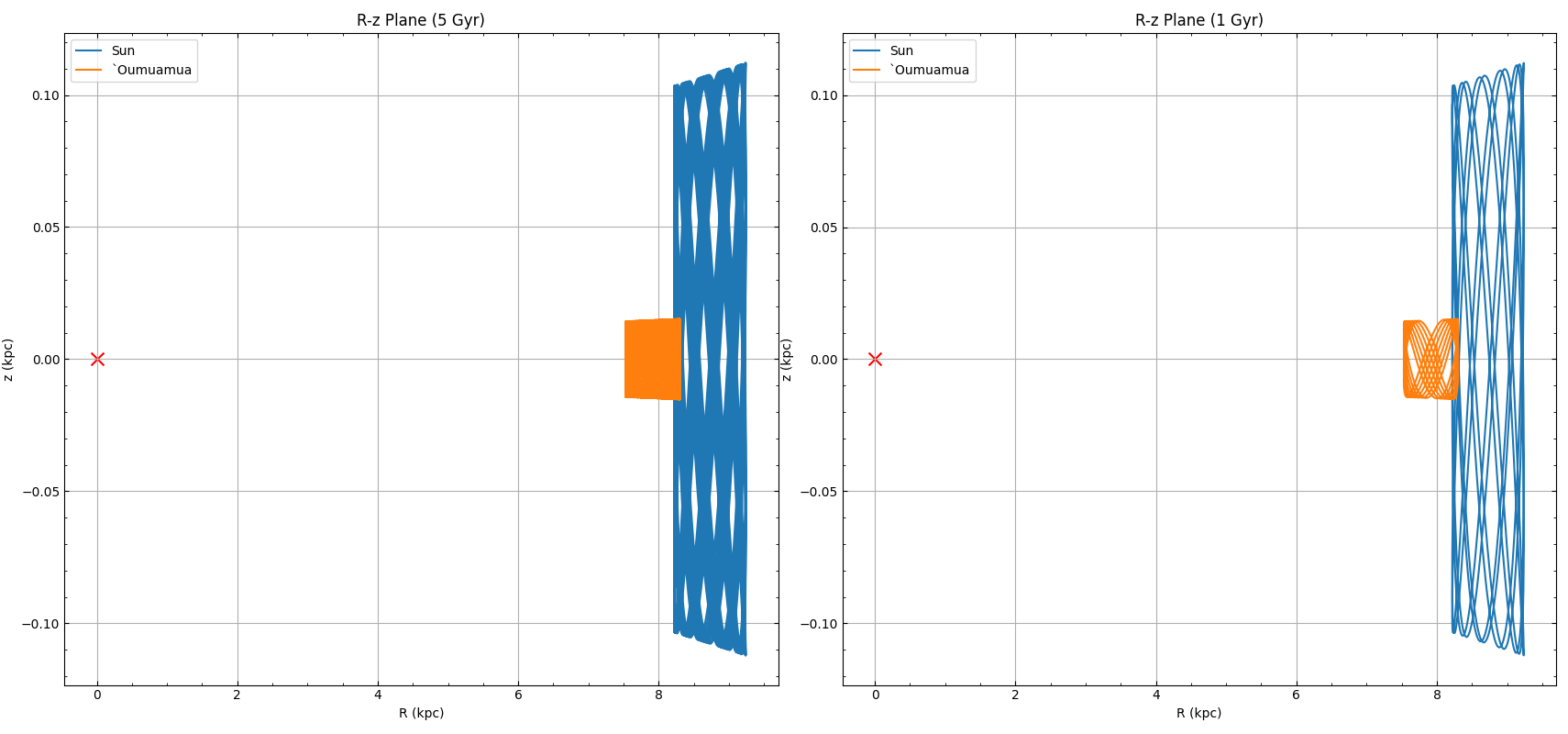}
    \caption{Trajectory of 1I/'Oumuamua in the $R-z$ plane of the Milky-Way disk. Left panel shows 5 Gyr integration, right panel shows 1 Gyr integration.}
    \label{Fig:O2}
\end{figure}

The radial ($R$) and vertical ($z$) extent of 1I/`Oumuamua's trajectory relative to the Galactic plane are depicted in figure~\ref{Fig:O2} for both 5 Gyr (left panel) and 1 Gyr (right panel) integrations. As a result of its low vertical velocity relative to the LSR, the vertical excursion of 1I/`Oumuamua (orange) is a factor of $\sim 6$ smaller than that of the Sun (blue). 

\begin{figure}[!ht]
    \centering
    \includegraphics[width=0.5\textwidth]{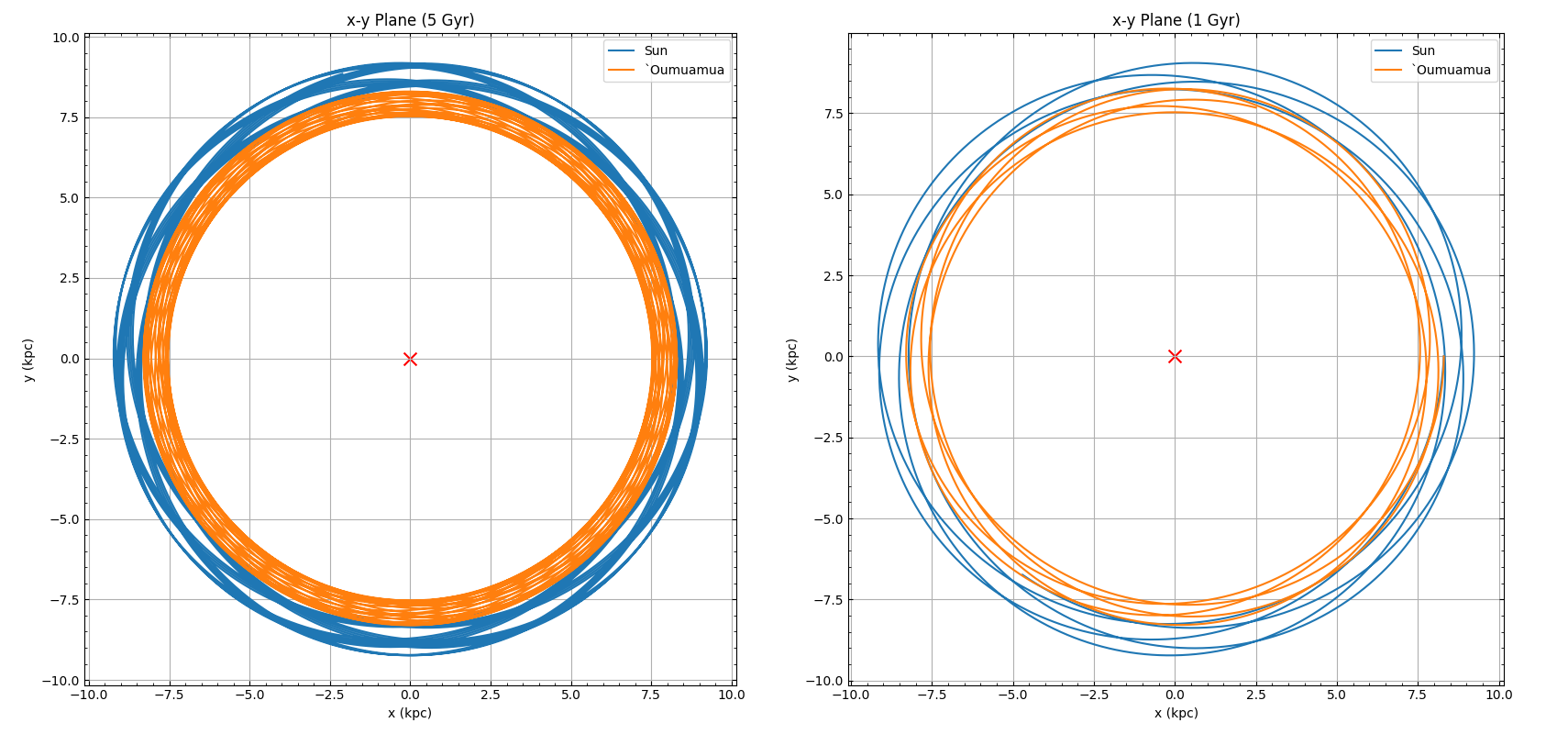}
    \caption{Trajectory of 1I/'Oumuamua in the $x-y$ plane of the Milky-Way disk. Left panel shows 5 Gyr integration, right panel shows 1 Gyr integration.}
    \label{Fig:O3}
\end{figure}

The motion in the Galactic plane ($x-y$) is shown in figure~\ref{Fig:O3}, comparing 5 Gyr and 1 Gyr integrations.

\subsubsection{Monte Carlo Analysis Results}

Applying the Monte Carlo methodology described in \S 2.1.1, the ensemble analysis yields a median maximum vertical excursion of $z_{\text{max}} = 0.016$ kpc with a 68\% confidence interval of $+0.003/-0.001$ kpc. The distribution spans from 0.015 to 0.038 kpc, reflecting combined measurement and systematic uncertainties. This tight distribution of small $z_{\text{max}}$ values strongly constrains 1I/`Oumuamua to have originated from a young stellar population with minimal dynamical heating.

\begin{figure}[!ht]
    \centering
    \includegraphics[width=0.5\textwidth]{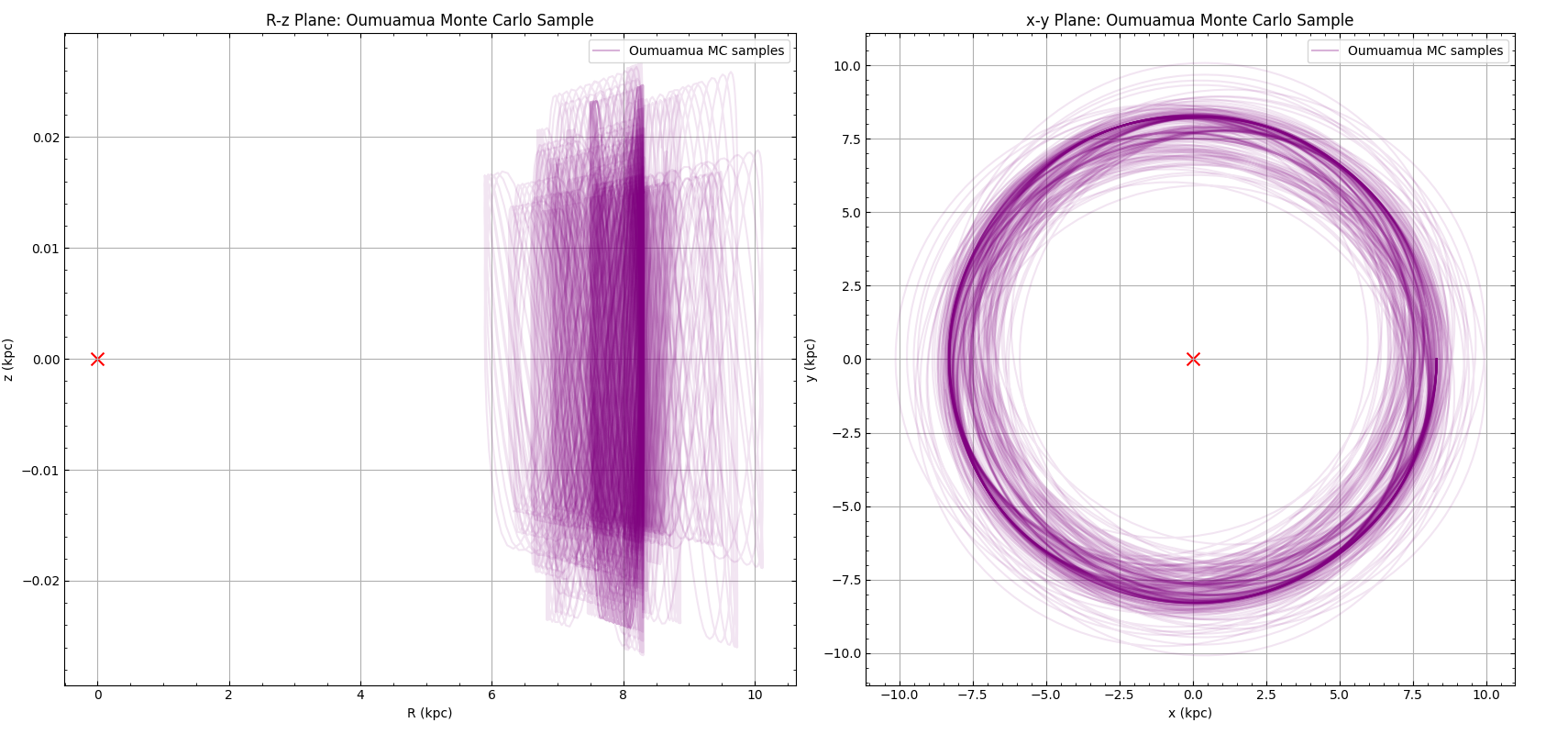}
    \caption{Monte Carlo ensemble of 1I/'Oumuamua trajectories in the $R-z$ plane. Transparent lines show 50 representative orbits from the uncertainty distribution, with the median trajectory highlighted.}
    \label{Fig:O4}
\end{figure}

\begin{figure}[!ht]
    \centering
    \includegraphics[width=0.45\textwidth]{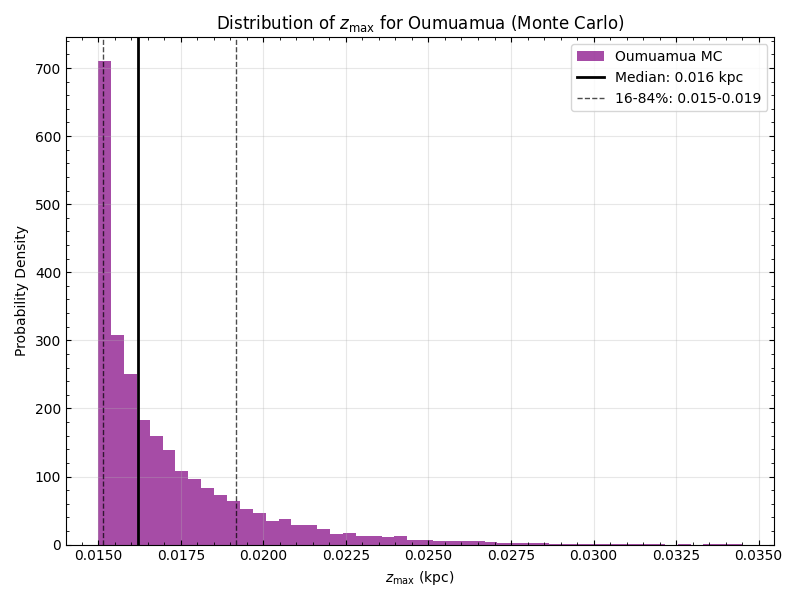}
    \caption{Distribution of maximum vertical excursions $z_{\text{max}}$ for 1I/'Oumuamua from Monte Carlo analysis. Vertical lines indicate median and 68\% confidence intervals.}
    \label{Fig:O5}
\end{figure}

\subsection{2I/Borisov}

\begin{figure}[!ht]
    \centering
    \includegraphics[width=0.45\textwidth]{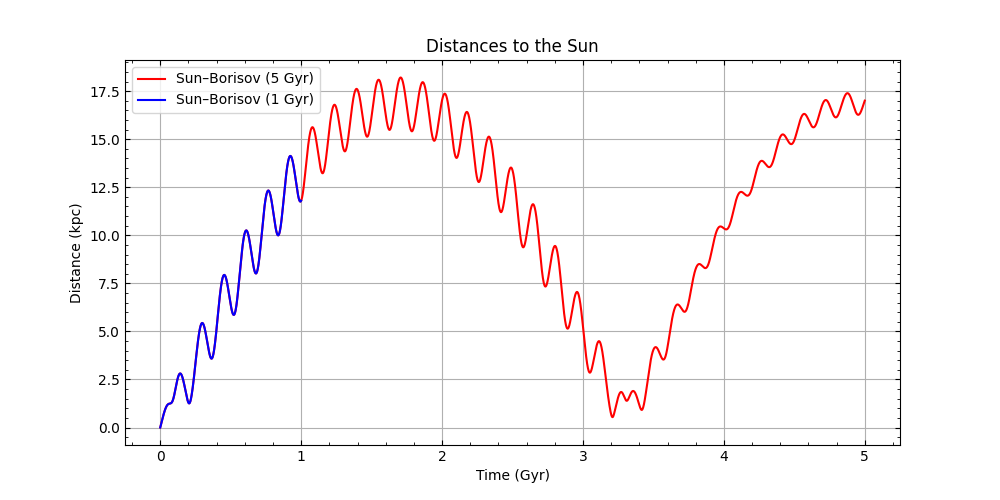}
    \caption{Distance of the comet 2I/Borisov from the Sun back in time with the present time represented by 0.}
    \label{Fig:B1}
\end{figure}

The past evolution of the distance of the interstellar comet 2I/Borisov from the Sun is shown in figure~\ref{Fig:B1}. The Sun-2I/Borisov separation follows a period of about 3.2 Gyr, with 2I/Borisov being on the other side of the Milky-Way disk relative to the Sun about 1.6 Gyr ago.

\begin{figure}[!ht]
    \centering
    \includegraphics[width=0.5\textwidth]{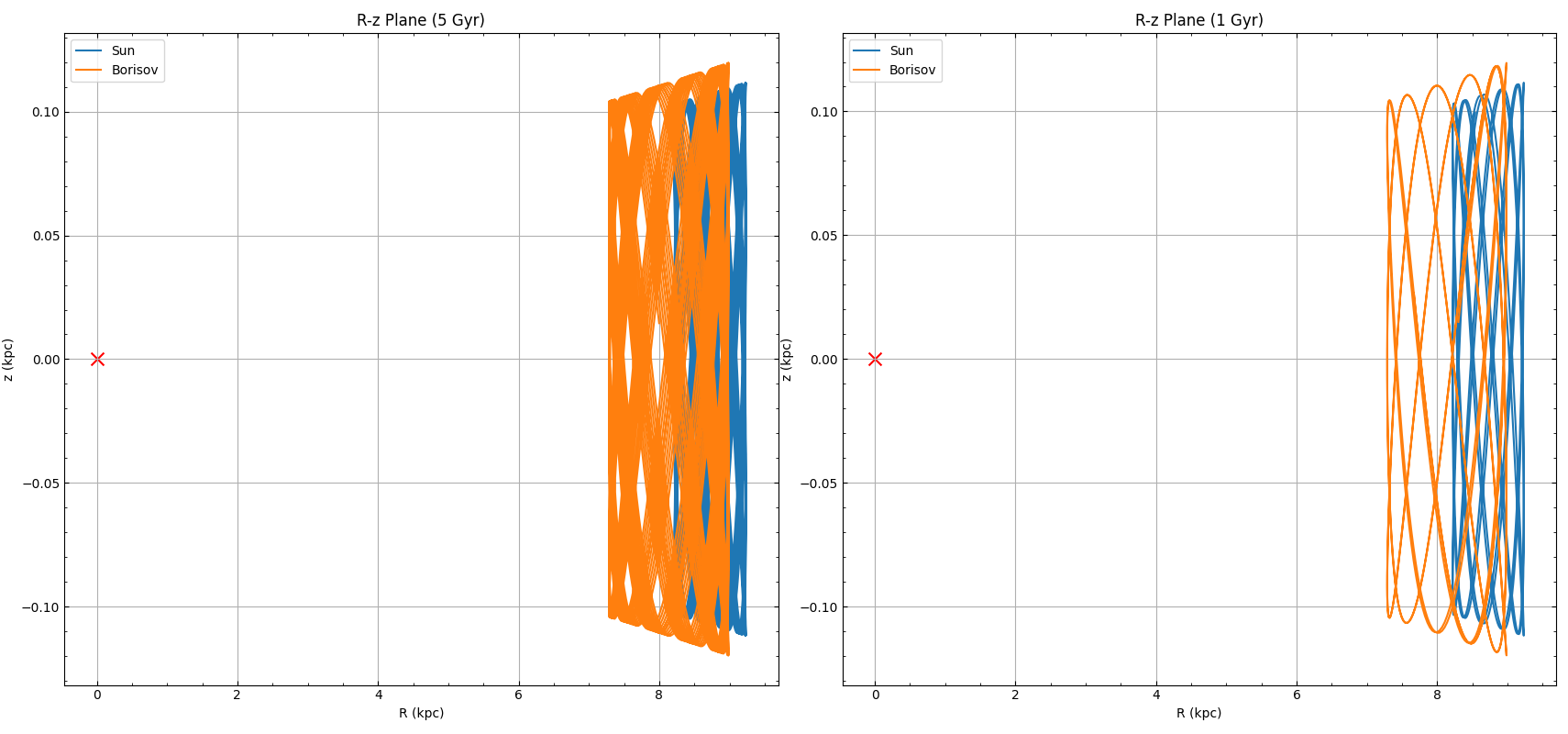}
    \caption{Trajectory of the comet 2I/Borisov in the $R-z$ plane of the Milky-Way disk. Left panel shows 5 Gyr integration, right panel shows 1 Gyr integration.}
    \label{Fig:B2}
\end{figure}

The radial ($R$) and vertical ($z$) extent of 2I/Borisov's trajectory relative to the Galactic plane are depicted in figure~\ref{Fig:B2} for both 5 Gyr (left panel) and 1 Gyr (right panel) integrations. The trajectories of 2I/Borisov (orange) resemble the corresponding ranges for the Sun (blue). 

\begin{figure}[!ht]
    \centering
    \includegraphics[width=0.5\textwidth]{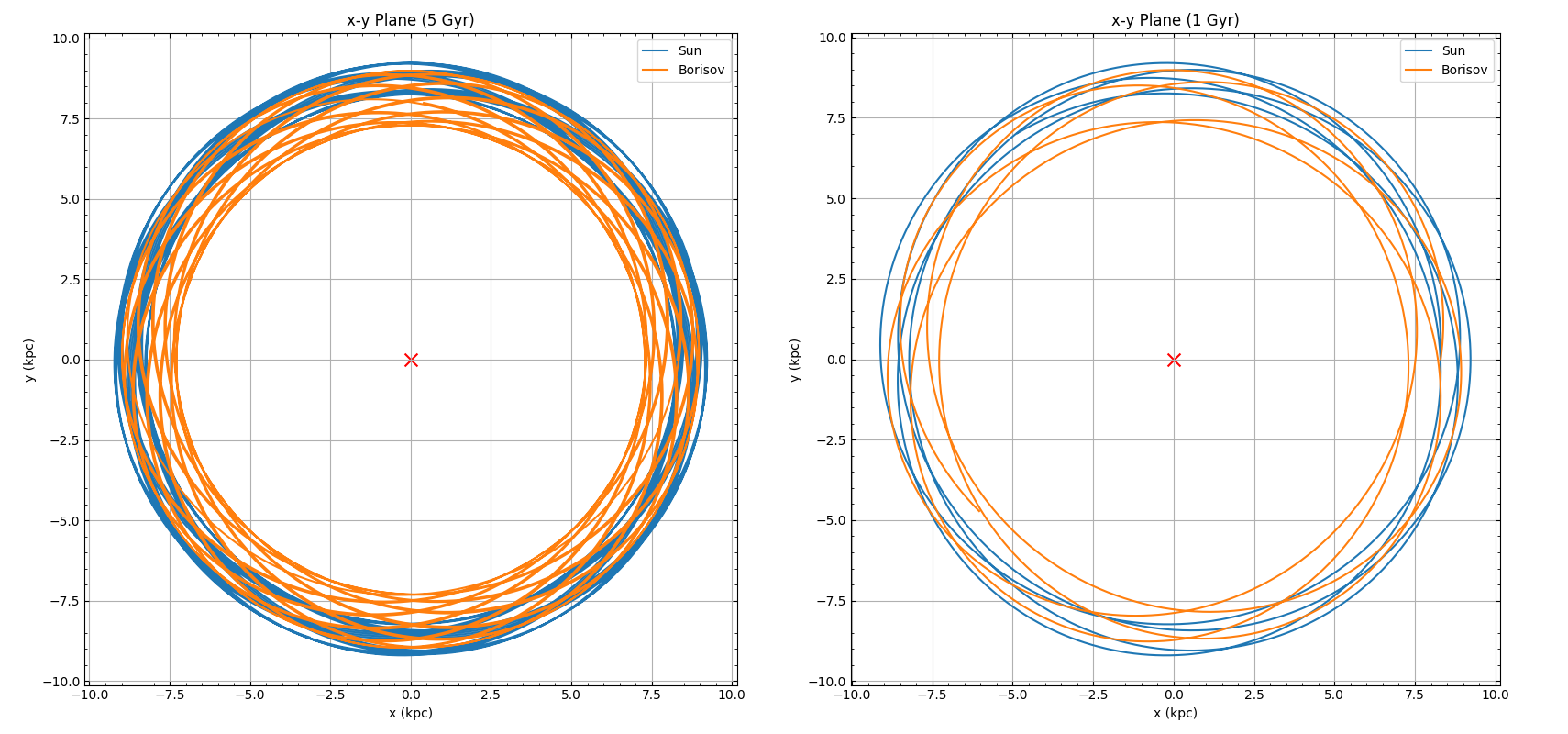}
    \caption{Trajectory of the comet 2I/Borisov in the $x-y$ plane of the Milky-Way disk. Left panel shows 5 Gyr integration, right panel shows 1 Gyr integration.}
    \label{Fig:B3}
\end{figure}

The motion in the Galactic plane ($x-y$) is shown in figure~\ref{Fig:B3}, again comparing 5 Gyr and 1 Gyr integrations.

\subsubsection{Monte Carlo Analysis Results}

The ensemble analysis yields a median maximum vertical excursion of $z_{\text{max}} = 0.121$ kpc with symmetric 68\% confidence intervals of $\pm 0.010$ kpc. The distribution ranges from 0.087 to 0.174 kpc. This intermediate $z_{\text{max}}$ value places 2I/Borisov in a regime consistent with moderately heated thin disk stars, confirming that the comet's vertical excursions are comparable to Solar-type orbits.

\begin{figure}[!ht]
    \centering
    \includegraphics[width=0.5\textwidth]{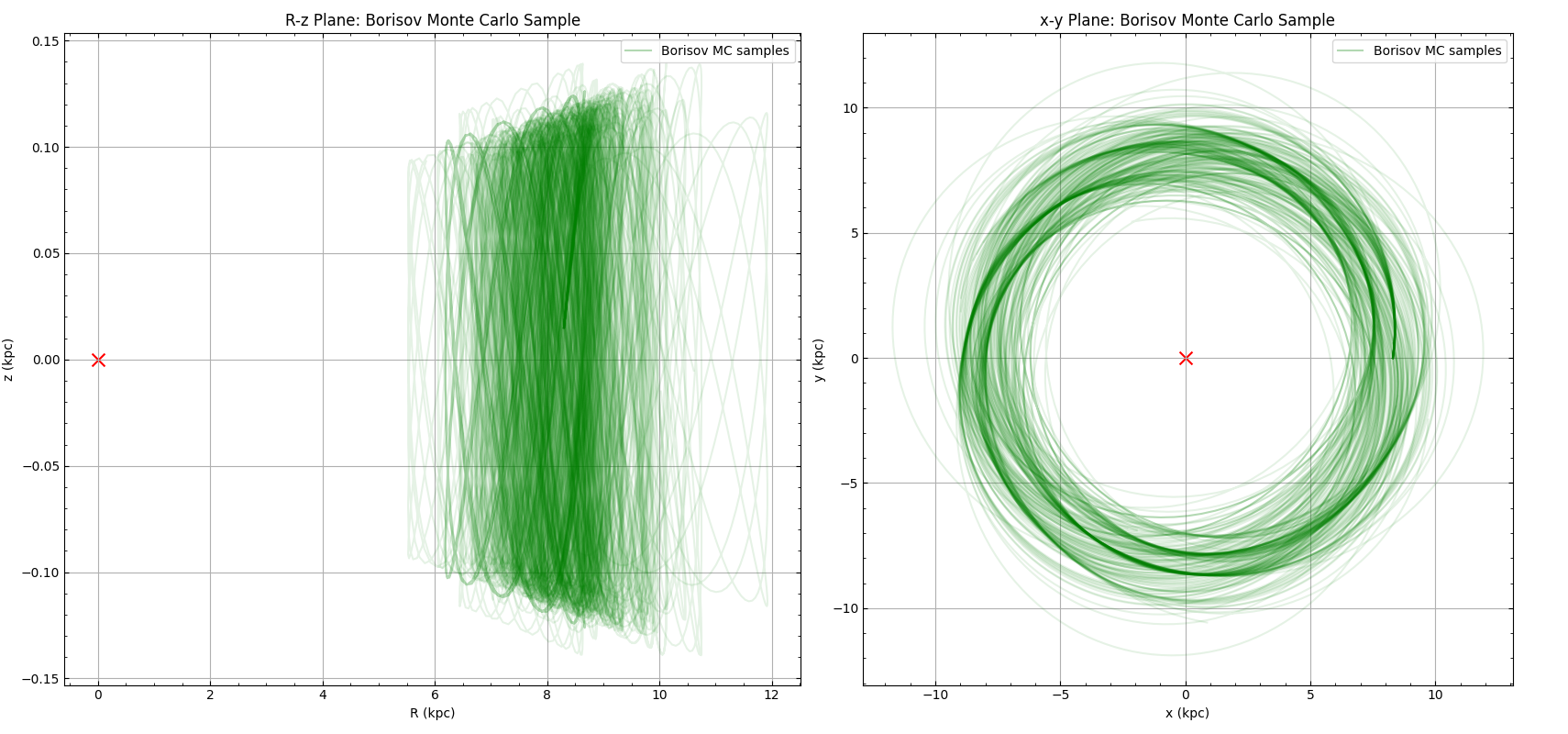}
    \caption{Monte Carlo ensemble of 2I/Borisov trajectories in the $R-z$ plane. Transparent lines show 50 representative orbits from the uncertainty distribution, with the median trajectory highlighted.}
    \label{Fig:B4}
\end{figure}

\begin{figure}[!ht]
    \centering
    \includegraphics[width=0.45\textwidth]{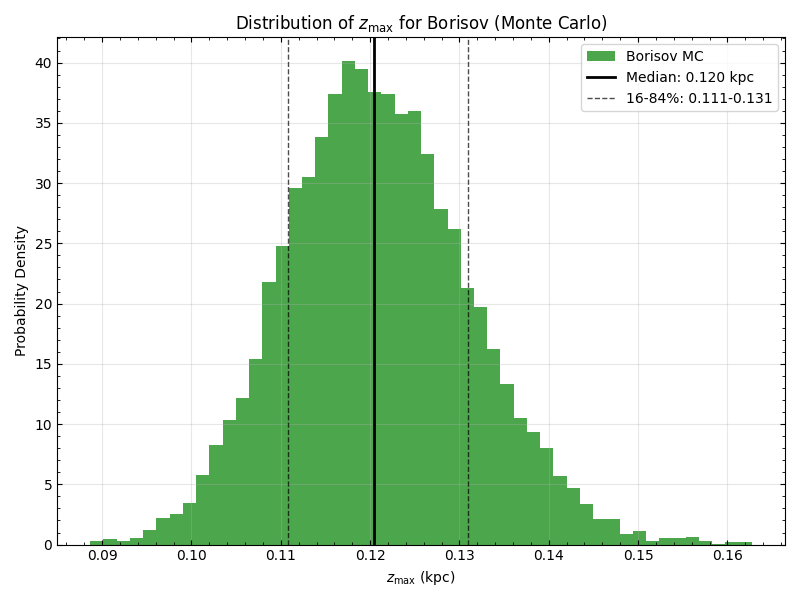}
    \caption{Distribution of maximum vertical excursions $z_{\text{max}}$ for 2I/Borisov from Monte Carlo analysis. Vertical lines indicate median and 68\% confidence intervals.}
    \label{Fig:B5}
\end{figure}

\subsection{3I/ATLAS}

\begin{figure}[!ht]
    \centering
    \includegraphics[width=0.45\textwidth]{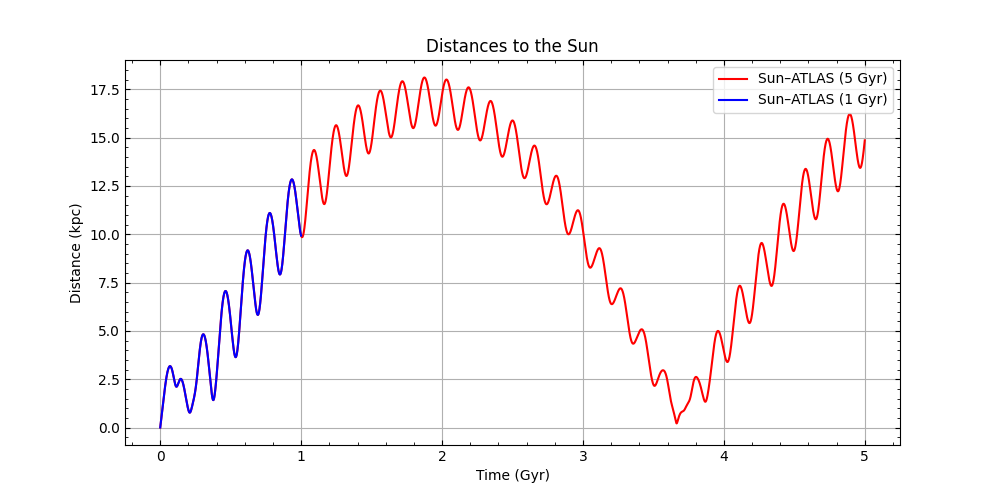}
    \caption{Distance of 3I/ATLAS from the Sun back in time.}
    \label{Fig:A1}
\end{figure}

The past evolution of the distance of the interstellar object 3I/ATLAS from the Sun is shown in figure~\ref{Fig:A1}. The 3I/ATLAS-Sun separation follows a period of about 3.63 Gyr, with 3I/ATLAS being on the other side of the Milky Way disk relative to the Sun about 1.815 Gyr ago.

\begin{figure}[!ht]
    \centering
    \includegraphics[width=0.5\textwidth]{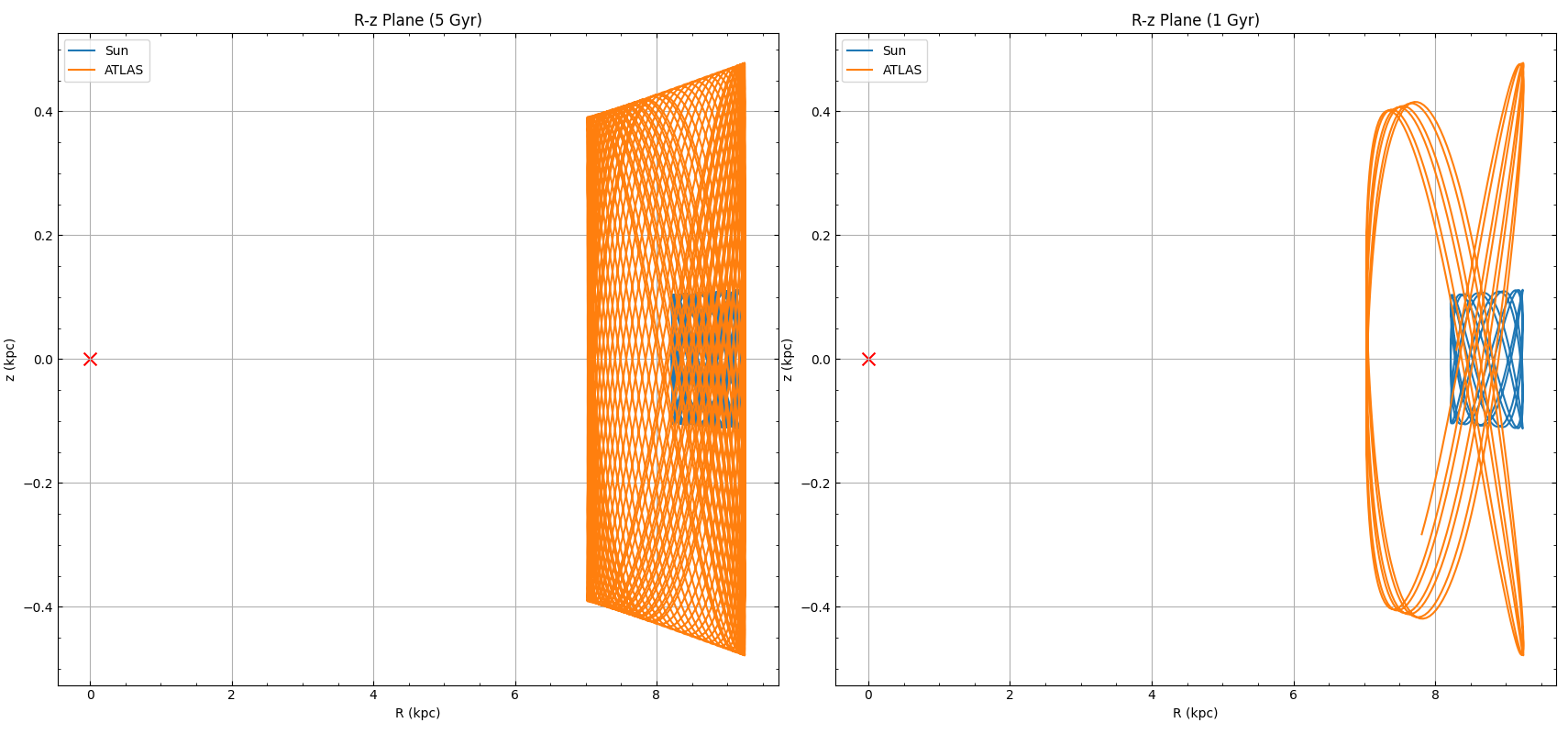}
    \caption{Trajectory of 3I/ATLAS in the $R-z$ plane of the Milky-Way disk. Left panel shows 5 Gyr integration, right panel shows 1 Gyr integration.}
    \label{Fig:A2}
\end{figure}

The radial ($R$) and vertical ($z$) extent of 3I/ATLAS's trajectory relative to the Galactic plane are depicted in figure~\ref{Fig:A2} for both 5 Gyr (left panel) and 1 Gyr (right panel) integrations. As a result of its vertical velocity, the vertical excursion of 3I/ATLAS (orange) is larger than that of the Sun (blue). 

\begin{figure}[!ht]
    \centering
    \includegraphics[width=0.5\textwidth]{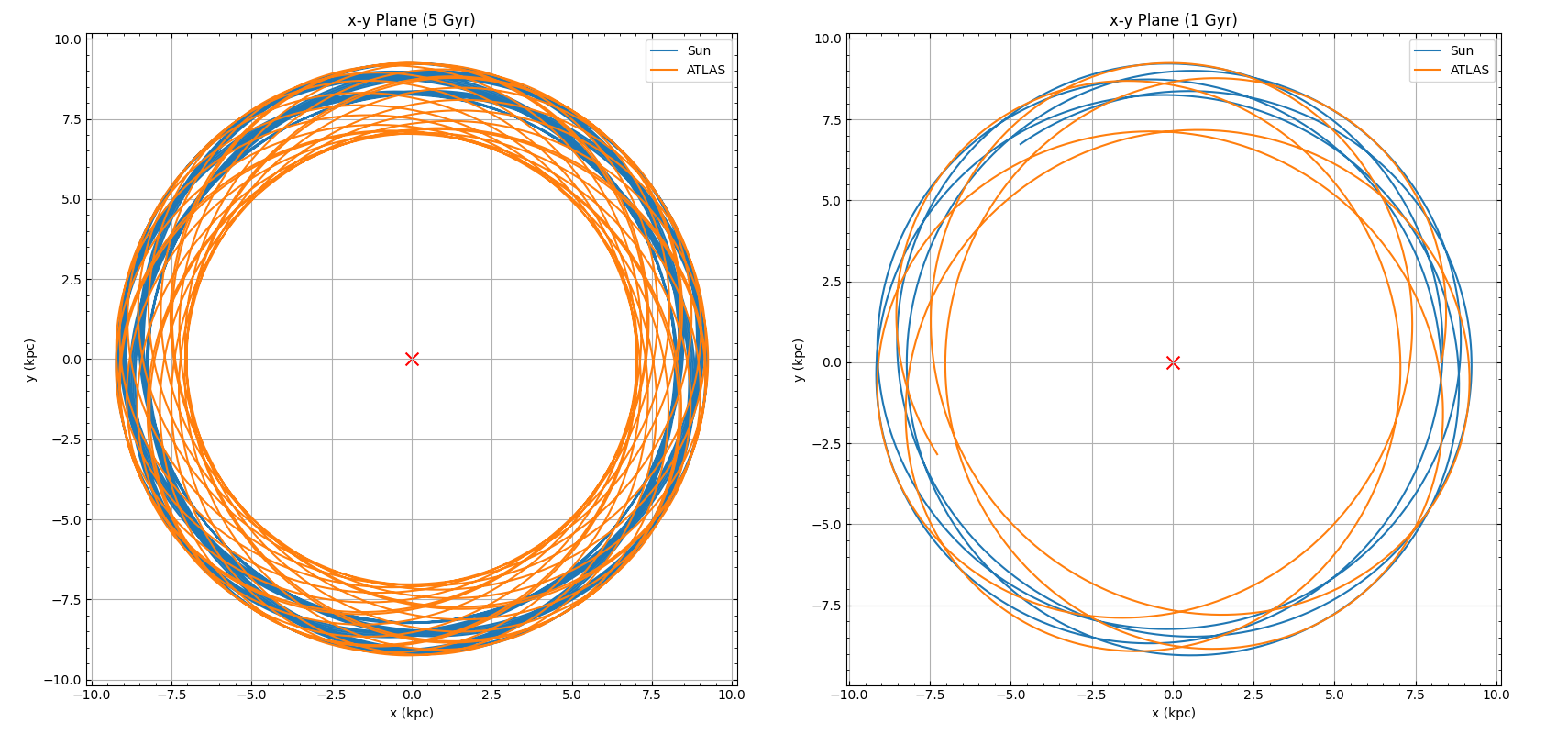}
    \caption{Trajectory of 3I/ATLAS in the $x-y$ plane of the Milky-Way disk. Left panel shows 5 Gyr integration, right panel shows 1 Gyr integration.}
    \label{Fig:A3}
\end{figure}

The motion in the Galactic plane ($x-y$) is shown in figure~\ref{Fig:A3}, comparing 5 Gyr and 1 Gyr integrations.

\subsubsection{Monte Carlo Analysis Results}

The ensemble analysis yields a median maximum vertical excursion of $z_{\text{max}} = 0.480$ kpc with a 68\% confidence interval of $+0.023/-0.017$ kpc. The distribution spans from 0.425 to 0.632 kpc. This large $z_{\text{max}}$ value, approaching half a kiloparsec, significantly exceeds the scale heights of young thin disk stars and Solar-age populations. The analysis confirms that 3I/ATLAS reaches vertical excursions characteristic of the thick disk component, indicating an origin from ancient stellar populations.

\begin{figure}[!ht]
    \centering
    \includegraphics[width=0.5\textwidth]{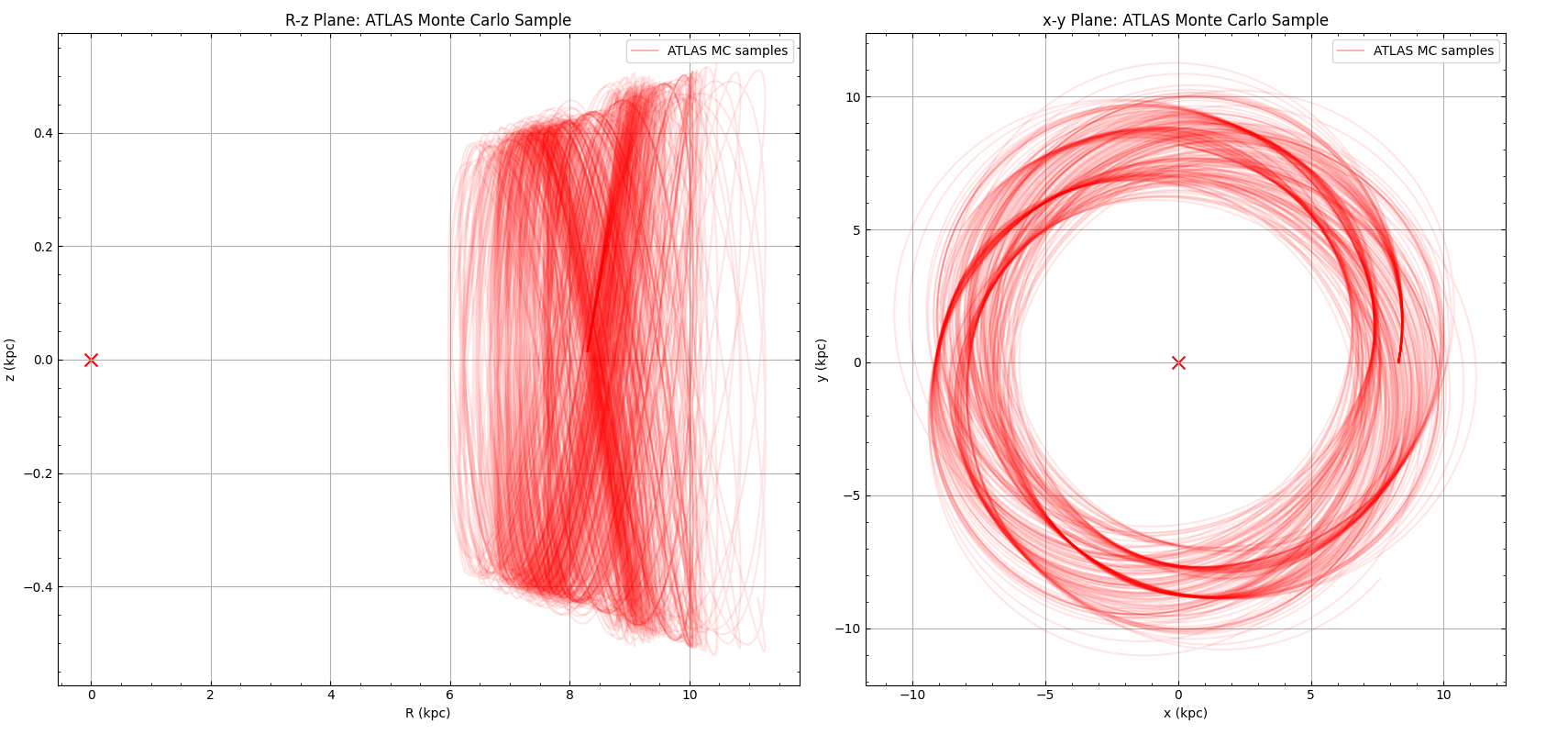}
    \caption{Monte Carlo ensemble of 3I/ATLAS trajectories in the $R-z$ plane. Transparent lines show 50 representative orbits from the uncertainty distribution, with the median trajectory highlighted.}
    \label{Fig:A4}
\end{figure}

\begin{figure}[!ht]
    \centering
    \includegraphics[width=0.45\textwidth]{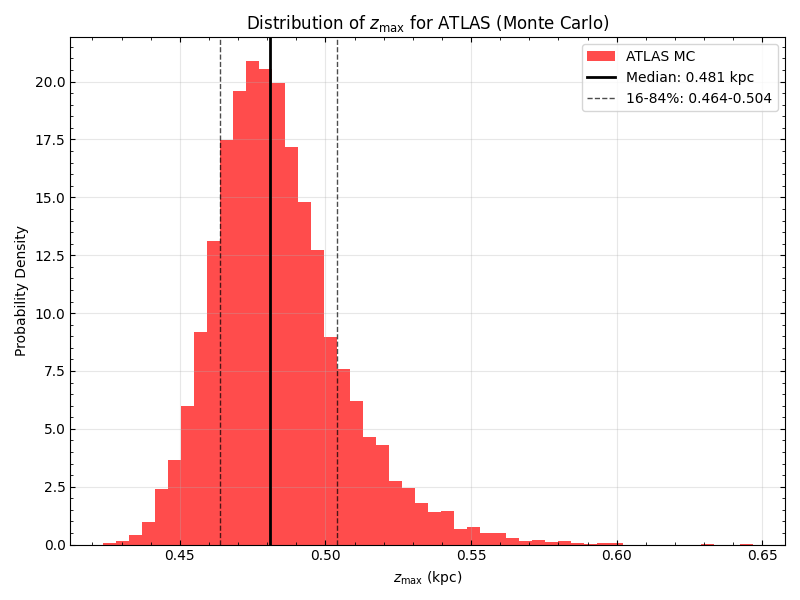}
    \caption{Distribution of maximum vertical excursions $z_{\text{max}}$ for 3I/ATLAS from Monte Carlo analysis. Vertical lines indicate median and 68\% confidence intervals.}
    \label{Fig:A5}
\end{figure}

\section{Probability Distribution of Age}

Next, we evaluate the probability distribution \( p(t) \) (with a unit normalization $\int p(t) dt=1$) for the age $t$ of the interstellar objects, assuming that they originated from stars. We base our analysis on  the extent of their vertical excursions, quantified by the maximum vertical ($z$) value that their orbits span,~\( z_{\text{max}} \), given that the scale-height of stars depends on their age. 

We present two complementary approaches for age inference: first, the traditional scale-height method based on stellar demographics (\S 3.1), and second, an improved Bayesian statistical framework that incorporates theoretical stellar kinematics (\S 3.2). The Bayesian approach addresses limitations of the demographic method by providing rigorous uncertainty quantification and incorporating physical models of dynamical heating processes.

\subsection{Traditional Scale-Height Method}

\subsubsection{Star Formation Rate (SFR) as a Function of Lookback Time}

\begin{figure}[!ht]
    \centering
    \includegraphics[width=0.45\textwidth]{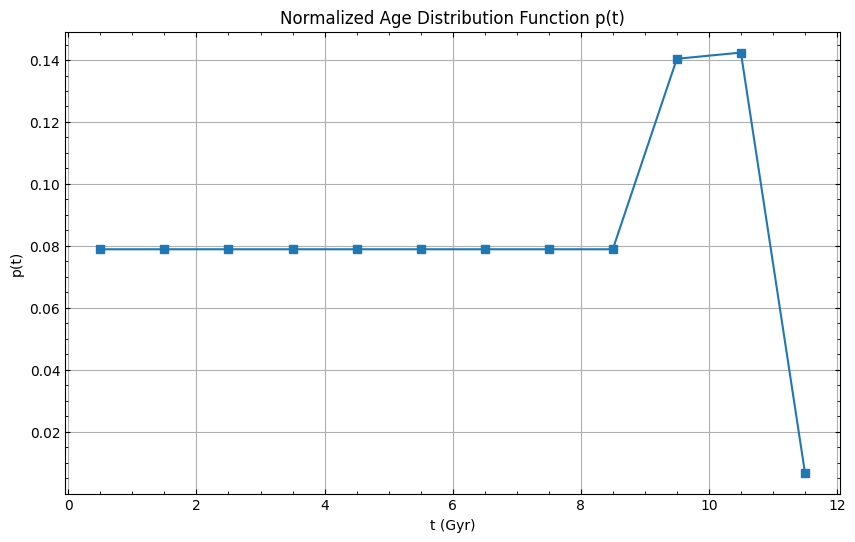}

    \caption{Probability distribution of age, \( p(t) \) in units of Gyr$^{-1}$, for the stars in the Milky Way disk, based on figure 9 in~\citet{2019ApJ...887..148F}.}
    \label{Fig:pt}
\end{figure}

We start from the empirically-calibrated star formation rate (SFR) history of the Milky-Way disk, based on the age distribution of white dwarfs~\citep{2019ApJ...887..148F}. The SFR is divided into two parts: a flat component from lookback time of 0 to 9 Gyr, plus a Gaussian component centered on 10 Gyr. For the flat component, we adopt a constant value of $\text{SFR}_{\text{linear}}$,
whereas the Gaussian part is described by, 
\begin{equation}
\text{SFR}_{\text{Gaussian}}(t) = \text{SFR}_{\text{peak}} \times \exp\left(-{\frac{1}{2}}\left(\frac{t - \text{t}_{\text{peak}}}{\sigma}\right)^2\right),
\end{equation}
where $\text{SFR}_{\text{peak}} = 3\text{SFR}_{\text{linear}}$, \(\text{t}_{\text{peak}} = 10\) Gyr, and \(\sigma = 0.5\) Gyr. Figure \ref{Fig:pt} shows the resulting age distribution of stars, plotted in 1 Gyr bins

The stellar mass formed in each temporal bin of width \(\Delta t=\)1 Gyr, is given by
\begin{equation}
\Delta M(t) = \text{SFR}_{\text{binned}}(t) \times \Delta t.
\end{equation}

The current number density of disk stars at the Galactic orbital radius of the Sun follow an exponential vertical distribution relative to the mid-plane, $\propto\exp\{-\vert z/h\vert\vert\}$, with a scale-height, $h(t)$, that depends on their age $t$. We adopt the observed $h(t)$ from \citet{2023MNRAS.522.1643C} and combine it with the star formation history described above to calibrate the fraction of stars with any given age. 

To quantify the age distribution of stars within the vertical excursion of each interstellar object, we determine the number of stars residing in each age bin up to the maximum $z$ value sampled by the object's trajectory, denoted by \(z_{\text{max}}\),
\begin{equation}
\Delta N(t) \propto \Delta M(t) \left(1-\exp\left(-\frac{z_{\text{max}}}{h(t)}\right)\right),
\end{equation}
where \( z_{\text{max}} \) is the maximum vertical height for the object.

To obtain a unit normalization of our probability distribution of ages, we find the total number of stars within the given \(z\) range as,
\begin{equation}
N_{\text{total}} = \sum_{i} \Delta N(t_i),
\end{equation}
where $N_{\text{total}}$ represents the total number of stars and $\Delta N(t)$ denotes the number of stars in the age bin centered at age \(t\).
The summation is performed over all age bins \(t_i\) given the constraint of $z_{\rm max}$.

The normalized age distribution function \( p(t) \) is then calculated from the relation,
\begin{equation}
p(t_i) = \frac{\Delta N(t_i)}{N_{\text{total}} \Delta t} .
\end{equation}

\subsubsection{Scale-Height Results for Individual Interstellar Objects}

\begin{figure}[!ht]
    \centering
    \includegraphics[width=0.35\textwidth]{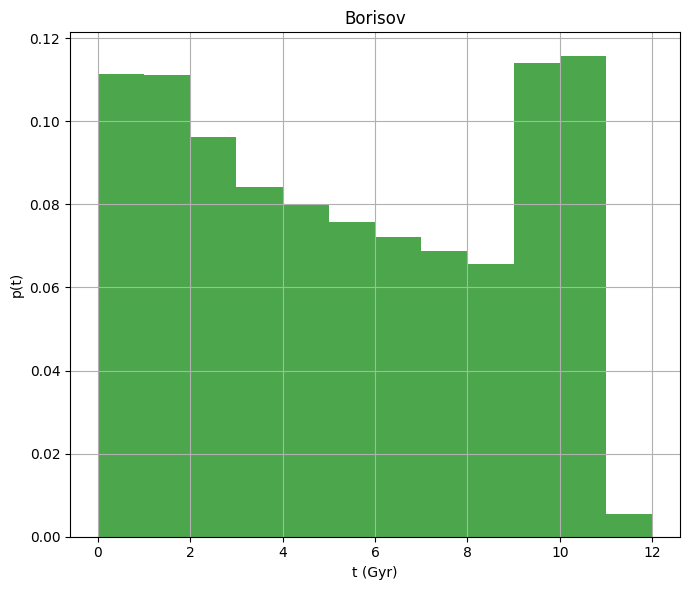}
    \caption{Probability distribution of age, \( p(t) \) in units of Gyr$^{-1}$, for the interstellar comet 2I/Borisov with \( z_{\text{max}} = 0.12 \) kpc.}
    \label{Fig:AgeB}
\end{figure}

Figure \ref{Fig:AgeB} shows the probability distribution of likely ages for the comet 2I/Borisov, which straddles the age of the Sun at $4.6$ Gyr.

\begin{figure}[!ht]
    \centering
\includegraphics[width=0.35\textwidth]{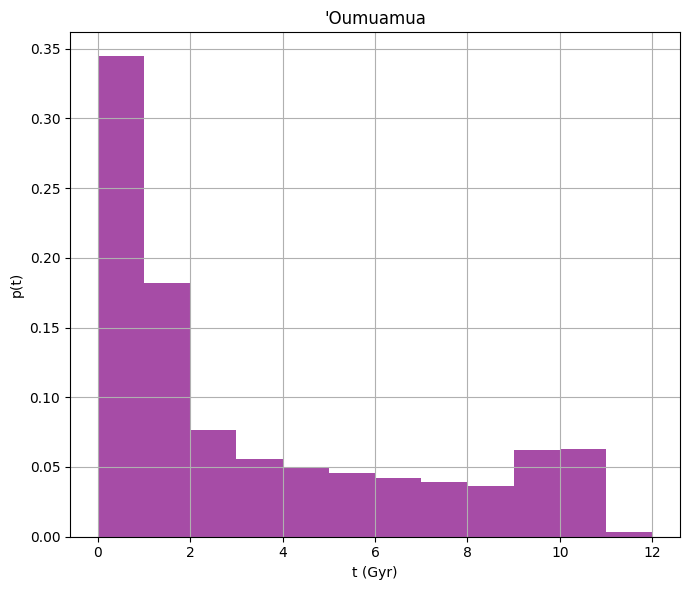}
    \caption{Probability distribution of age, \( p(t) \) in units of Gyr$^{-1}$, for 1I/`Oumuamua with \( z_{\text{max}} = 0.015 \) kpc.}.
        \label{Fig:AgeO}
        \end{figure}

In contrast, 1I/`Oumuamua is much younger with a likely age of 1-2 Gyr, based on its much smaller value of $z_{\rm max}$, as shown in Figure~\ref{Fig:AgeO}.

\begin{figure}[!ht]
    \centering
    \includegraphics[width=0.35\textwidth]{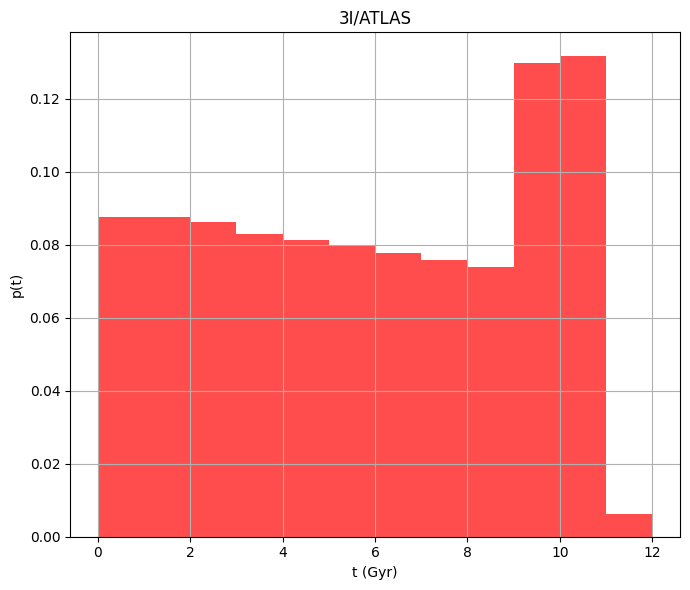}
    \caption{Probability distribution of age, \( p(t) \) in units of Gyr$^{-1}$, for 3I/ATLAS with \( z_{\text{max}} = \text{0.475} \) kpc.}.
        \label{Fig:AgeA}
        \end{figure}

Finally, 3I/ATLAS may follow a probability distribution similar to the entire Galactic disk population because of its potentially large value of $z_{\rm max}$, as shown in Figure~\ref{Fig:AgeA}.

\subsection{Bayesian Statistical Framework}

To address the limitations of the demographic approach and provide rigorous uncertainty quantification, we implement a Bayesian statistical framework that combines the Monte Carlo orbital results with theoretical models of stellar kinematics. This methodology provides quantitative age estimates with proper credible intervals while incorporating the physical processes governing stellar velocity evolution.

\subsubsection{Theoretical Foundation}

The Bayesian approach relies on the fundamental correlation between stellar age and vertical velocity dispersion in the galactic disk. Young stellar populations exhibit small velocity dispersions due to limited dynamical heating, while old populations show large dispersions from billions of years of gravitational scattering by giant molecular clouds, spiral arms, and dark matter substructure.

Our posterior probability for stellar age $\tau$ given observed maximum vertical height $z_{\text{max}}$ follows Bayes' theorem:
\begin{equation}
P(\tau|z_{\text{max}}) \propto P(z_{\text{max}}|\tau) \times \text{SFR}(\tau)
\end{equation}
where $P(z_{\text{max}}|\tau)$ is the likelihood function and $\text{SFR}(\tau)$ represents the star formation rate prior encoding when stars formed in galactic history.

\subsubsection{Likelihood Function}

We adopt a Rayleigh-like distribution for the probability of observing a maximum vertical height $z_{\text{max}}$ for a star of age $\tau$:
\begin{equation}
P(z_{\text{max}}|\tau) = \frac{\nu^2 z_{\text{max}}}{\sigma_z(\tau)^2} \exp\left(-\frac{\nu^2 z_{\text{max}}^2}{2\sigma_z(\tau)^2}\right)
\end{equation}
This functional form emerges naturally from assuming stellar velocities follow three-dimensional Gaussian distributions combined with harmonic oscillator dynamics in the galactic gravitational potential~\citep{2008gady.book.....B}.

The vertical frequency squared $\nu^2 = 1200.0$ (km s$^{-1}$)$^2$ kpc$^{-2}$ represents the second derivative of the galactic potential at the midplane, determining the restoring force strength for vertical oscillations. This value is calibrated from the galactic potential model and ensures consistency with observed stellar kinematics in the solar neighborhood.

\subsubsection{Age-Dependent Velocity Dispersion}

The key formula is the physically motivated velocity dispersion model $\sigma_z(\tau)$ that captures the distinct kinematic properties of thin and thick galactic disk populations:

For $\tau \leq 7$ Gyr (thin disk):
\begin{equation}
\sigma_z(\tau) = 2.0 \times \left(\frac{\tau}{1 \text{ Gyr}}\right)^{0.4} \text{ km s}^{-1}
\end{equation}

For $\tau > 7$ Gyr (thick disk):
\begin{equation}
\sigma_z(\tau) = 40.0 \text{ km s}^{-1}
\end{equation}

The power-law component reflects gradual dynamical heating from gravitational scattering by giant molecular clouds and stellar clusters. \citet{1992PASJ...44..601K} demonstrated through epicycle orbit integration that scattering by giant molecular clouds produces velocity evolution $\sigma(\tau) \propto \tau^x$ with $x \approx 0.25$ for all three velocity components, reaching $\sigma_z \sim 10$ km s$^{-1}$ after 10 Gyr for typical GMC masses of $\sim 10^6$ M$_\odot$. Our adopted exponent $\beta = 0.4$ represents a compromise between these theoretical predictions and observational constraints~\citep{2009MNRAS.397.1286A}. 

\citet{2002MNRAS.330..707K} identified additional heating mechanisms where unbound stellar clusters expand at super-virial speeds after gas expulsion, producing high-velocity tails that help populate the dispersion wing and contribute to steeper age-velocity relations. The normalization $\sigma_1 = 2.0$ km s$^{-1}$ at 1 Gyr is consistent with observations by \citet{1999A&A...350..434A}, who showed that young disk stars start with extremely small vertical dispersions ($\sigma_W \sim 2-5$ km s$^{-1}$) and that purely secular heating cannot reach 30 km s$^{-1}$ without invoking additional mechanisms.

The thick disk component reflects the observed kinematic dichotomy in galactic stellar populations, where stars older than 7 Gyr exhibit systematically higher velocity dispersions due to early violent galactic evolution and different formation processes. The transition to $\sigma_z = 40.0$ km s$^{-1}$ at 7 Gyr is motivated by LAMOST-Gaia observations of age-velocity dispersion relations in the galactic disk~\citep{2018MNRAS.475.1093Y}.

\subsubsection{Star Formation Rate Prior}

We implement an exponential decline $\text{SFR}(\tau) = \exp(-\tau/8.0 \text{ Gyr})$ representing the general trend of decreasing star formation over cosmic time. The 8 Gyr decay timescale is consistent with observational constraints from stellar archaeology studies~\citep{2000ApJ...531..965R,2004A&A...423..517R} and captures the decline in galactic star formation activity since the peak epoch around $z \sim 1-2$.

\subsubsection{Computational Implementation and Results}

The analysis employs a high-resolution age grid from 1 Myr to 12 Gyr with 1 Myr increments, providing sufficient temporal resolution to capture sharp features in the posterior distributions. For each age bin, we evaluate the likelihood function using the median $z_{\text{max}}$ values from our Monte Carlo orbital analysis:

\begin{itemize}
\item 1I/`Oumuamua: $z_{\text{max}} = 0.016$ kpc
\item 2I/Borisov: $z_{\text{max}} = 0.121$ kpc  
\item 3I/ATLAS: $z_{\text{max}} = 0.480$ kpc
\end{itemize}

The posterior distribution is computed as the normalized product of likelihood and prior, with statistical inference based on robust percentile methods. We report median ages as point estimates and derive credible intervals from the 16th/84th percentiles (68\% confidence) and 2.5th/97.5th percentiles (95\% confidence).

\subsubsection{Bayesian Age Inference Results}

The Bayesian analysis yields quantitative age estimates that demonstrate clear physical trends:

\textbf{1I/`Oumuamua (young stellar system):} Median age of 1.0 Gyr with 68\% credible interval of 0.1-4.1 Gyr. The extremely small $z_{\text{max}}$ value strongly constrains the origin to young stellar populations with minimal dynamical heating.

\textbf{2I/Borisov (intermediate-age population):} Median age of 3.8 Gyr with 68\% credible interval of 1.8-5.9 Gyr. The moderate $z_{\text{max}}$ indicates origin from thin disk stars with intermediate levels of dynamical heating.

\textbf{3I/ATLAS (old thick-disk source):} Median age of 9.6 Gyr with 68\% credible interval of 7.8-10.3 Gyr. The large $z_{\text{max}}$ requires high velocity dispersions characteristic of ancient thick disk populations.

These results demonstrate the power of Bayesian age discrimination, where smaller vertical excursions correspond systematically to younger stellar origins and larger excursions indicate older populations. The methodology provides robust statistical uncertainties that properly account for both observational limitations and theoretical model uncertainties.

\begin{figure}[!ht]
    \centering
    \includegraphics[width=0.45\textwidth]{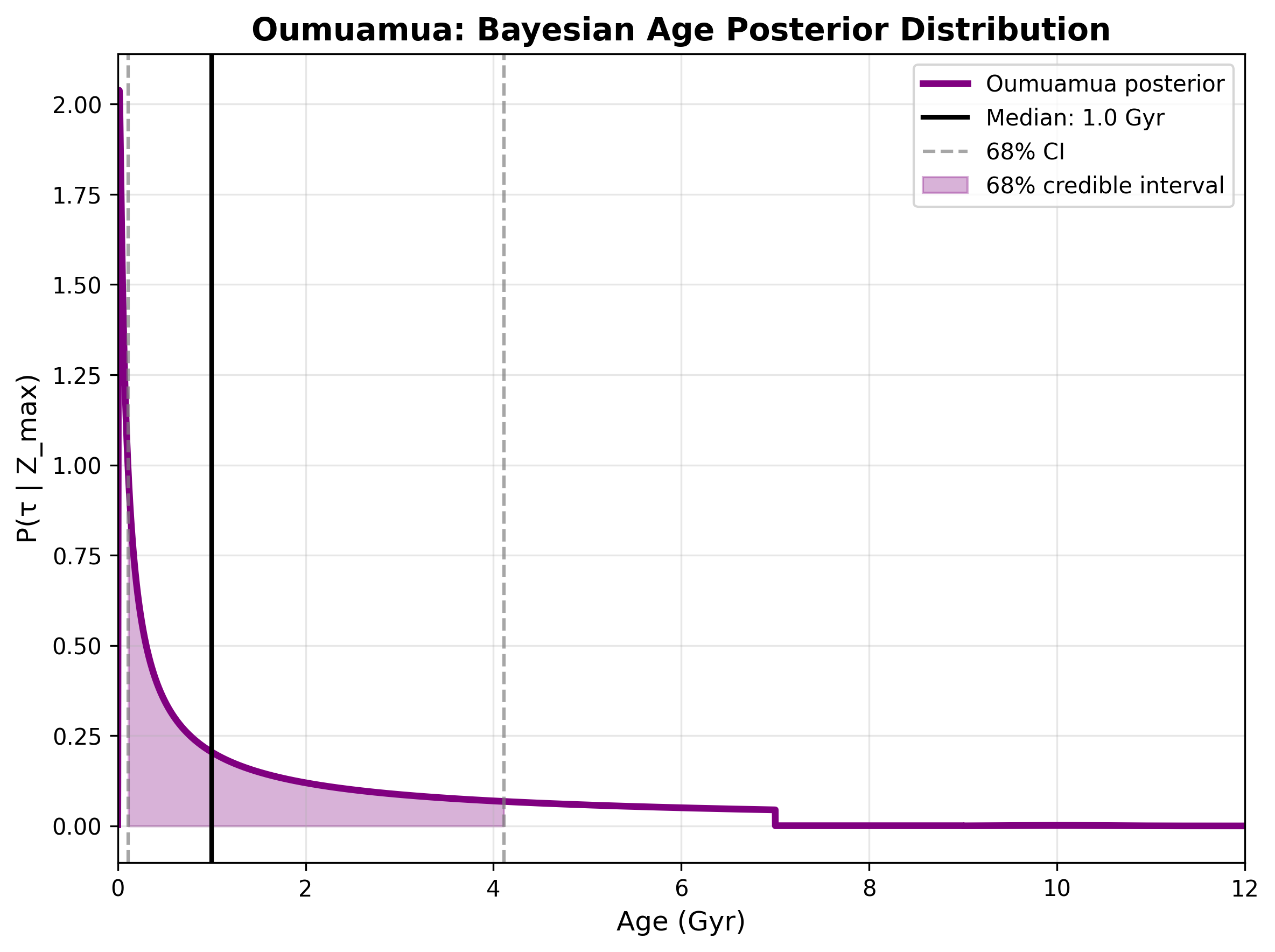}
    \caption{Bayesian age posterior for 1I/'Oumuamua. The posterior distribution shows a median age of 1.0 Gyr with 68\% credible interval of [0.1, 4.1] Gyr, based on $z_{\text{max}} = 0.016$ kpc. The young age and tight distribution indicate origin from a recently formed stellar system in the thin galactic disk with minimal dynamical heating.}
    \label{Fig:BayesianOumuamua}
\end{figure}

\begin{figure}[!ht]
    \centering
    \includegraphics[width=0.45\textwidth]{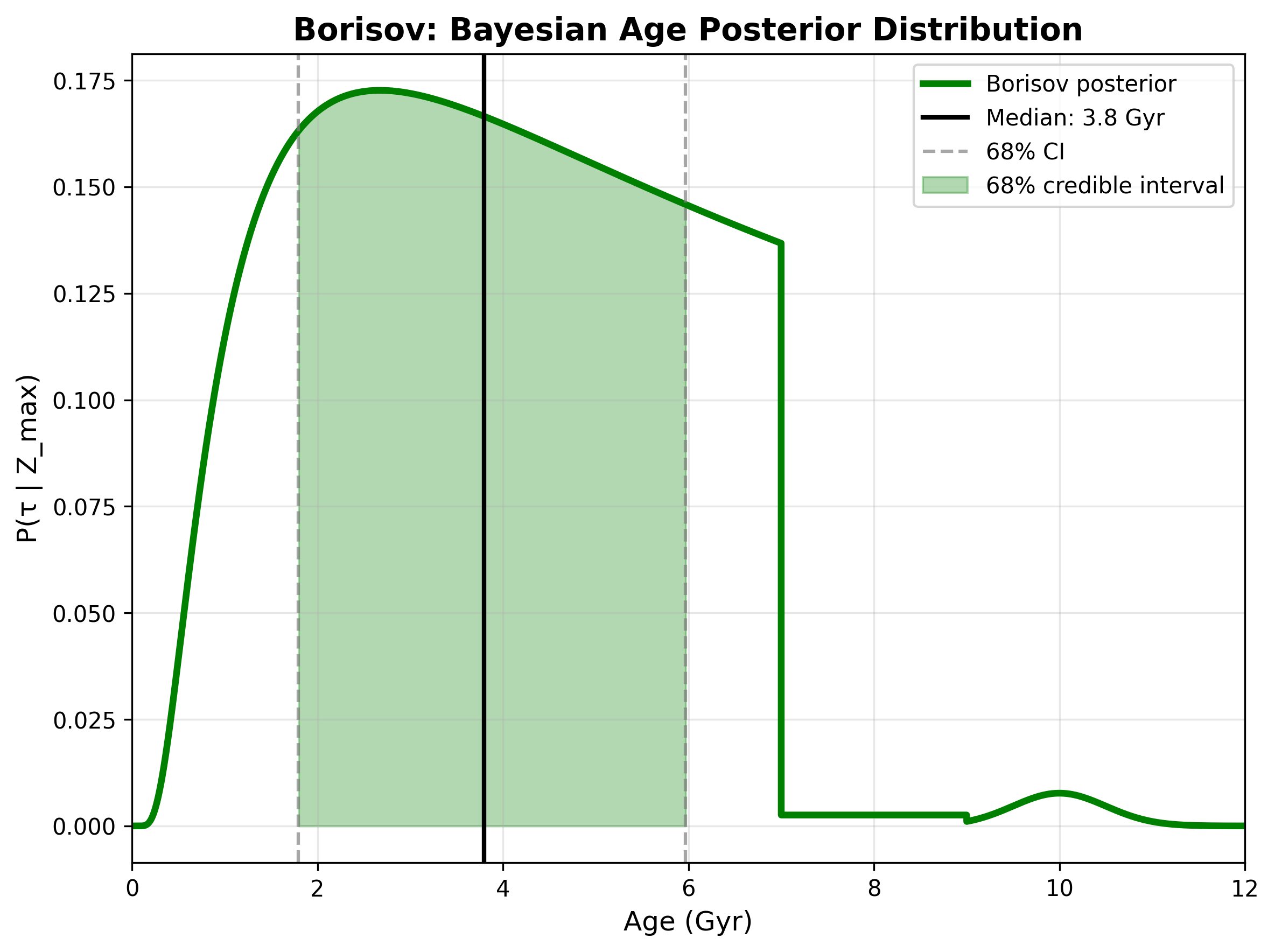}
    \caption{Bayesian age posterior for 2I/Borisov. The posterior distribution shows a median age of 3.8 Gyr with 68\% credible interval of [1.8, 5.9] Gyr, based on $z_{\text{max}} = 0.121$ kpc. The intermediate age suggests origin from a moderately heated thin disk stellar population comparable to Solar-type orbits.}
    \label{Fig:BayesianBorisov}
\end{figure}

\begin{figure}[!ht]
    \centering
    \includegraphics[width=0.45\textwidth]{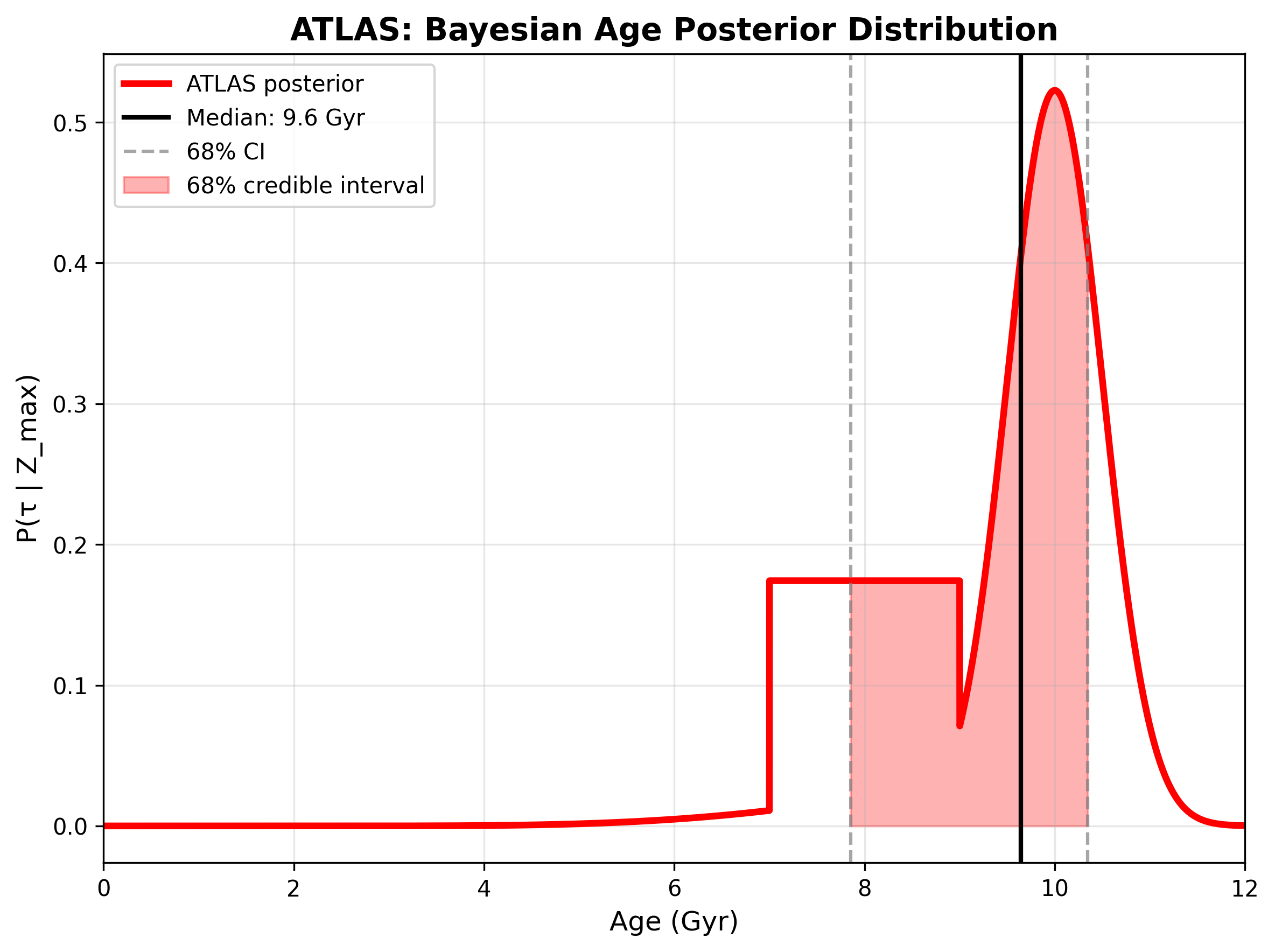}
    \caption{Bayesian age posterior for 3I/ATLAS. The posterior distribution shows a median age of 9.6 Gyr with 68\% credible interval of [7.8, 10.3] Gyr, based on $z_{\text{max}} = 0.480$ kpc. The ancient age indicates origin from the thick disk component formed during early violent galactic evolution.}
    \label{Fig:BayesianATLAS}
\end{figure}

\begin{figure}[!ht]
    \centering
    \includegraphics[width=0.5\textwidth]{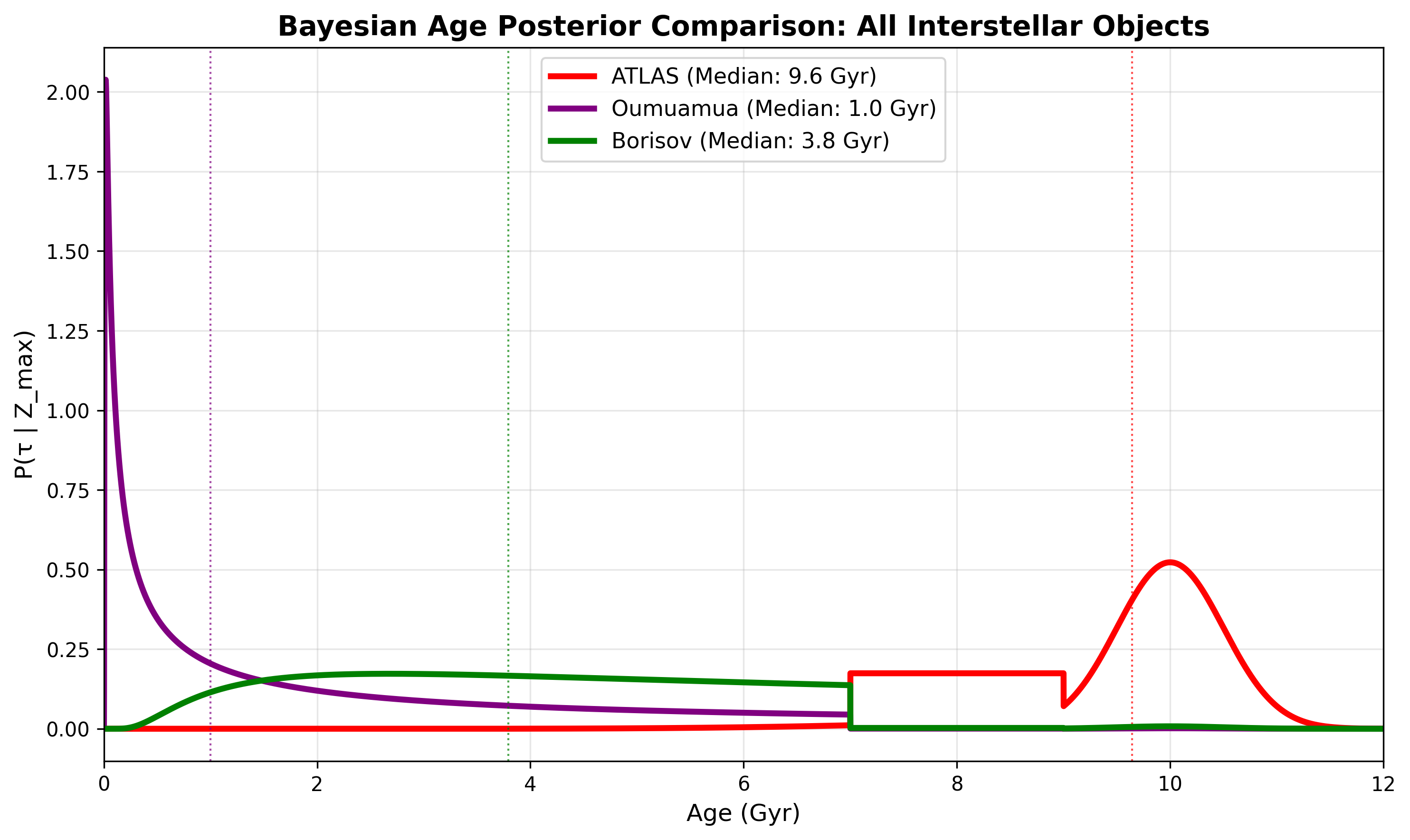}
    \caption{Comparison of Bayesian age posteriors for all three interstellar objects. 1I/'Oumuamua (purple) originated from a young thin disk stellar population (1.0 Gyr), 2I/Borisov (green) from an intermediate-age population (3.8 Gyr), and 3I/ATLAS (red) from an ancient thick disk source (9.6 Gyr). The diversity of ages demonstrates multiple formation pathways operating throughout galactic history.}
    \label{Fig:BayesianComparison}
\end{figure}

\subsection{Comparison with Previous Studies}

Our Bayesian age inference results can be directly compared to previous investigations of interstellar object origins and ages, providing important context for our findings and highlighting methodological advances.

\subsubsection{1I/'Oumuamua Age Constraints}

\citet{2018MNRAS.480.4903A} performed kinematic age analysis for 1I/'Oumuamua using Bayesian inference based on the Geneva-Copenhagen Survey. Their approach assumed ISOs initially share their parental star's Galactic orbit and experience subsequent dynamical heating similar to field stars. They obtained an age estimate for 'Oumuamua in the range 0.01--1.87 Gyr, with a preferred age between 0.20--0.45 Gyr, depending on their adopted age-velocity dispersion relation model. Our inferred median age (1.0 Gyr, 68\% credible interval: 0.1--4.1 Gyr) is broadly consistent with their results, while offering improved uncertainty quantification through explicit error propagation and more detailed age-velocity dispersion modeling that incorporates both thin and thick disk populations.

\citet{2020AJ....159..147H} highlighted fundamental limitations in tracing precise Galactic trajectories for ISOs, emphasizing that backward integrations lose reliability beyond $\sim$10 Myr due to disk heating effects. They concluded that 1I/'Oumuamua's relatively low velocity dispersion ($\sim$3--10 km s$^{-1}$ relative to the Local Standard of Rest) implies a likely young origin (less than 100 Myr). While our Bayesian age estimate for 1I/'Oumuamua is slightly older than their inferred young age, our credible interval encompasses their conclusion, and the apparent discrepancy reflects the large uncertainties inherent to orbital reconstructions and differences in Galactic dynamical modeling assumptions.

\citet{2021ApJ...917...20H} conducted detailed backward orbital integrations and Monte Carlo simulations to trace the Galactic trajectory of 1I/'Oumuamua. They concluded that 'Oumuamua likely originated from the Carina or Columba young stellar associations approximately 30--35 Myr ago, based on dynamical arguments comparing orbital parameters such as maximum vertical excursion ($z_{\rm max}$) and radial position within the Galaxy. Their simulations highlighted significant alignment with these young stellar groups, providing strong evidence for a recent (tens of Myr) ejection event. Unlike our broader Bayesian age analysis, which provides a wider credible interval (median 1.0 Gyr, 68\% CI: 0.1--4.1 Gyr), Hsieh et al.'s approach strongly favors a younger origin ($\sim$30 Myr). This discrepancy arises primarily due to methodological differences: Hsieh et al. used detailed backtracking and spatial intersection with young stellar associations, whereas our analysis integrates uncertainties over a longer Galactic dynamical timescale, considering the full stellar velocity dispersion profile. Despite these differences, both studies support a relatively young dynamical age for 1I/'Oumuamua, though our probabilistic method naturally yields a broader range reflecting observational uncertainties and dynamical heating processes over extended periods.

\subsubsection{Trajectory Analysis and Stellar Origins}

\citet{2018AJ....156..205B} explored potential stellar origin systems for both 1I/'Oumuamua and the broader population of interstellar objects using precise astrometry. They emphasized the fundamental challenges in tracing ISOs to specific stellar origins due to uncertainties in both object trajectories and stellar motions over long timescales. Our probabilistic approach complements their findings by quantifying the broader age demographics that such objects represent, rather than attempting to identify specific stellar origins, thus providing a statistical framework for understanding ISO populations.

The star formation history adopted in our Bayesian prior draws from \citet{2019ApJ...887..148F}, who reconstructed the Milky Way's star formation history using white dwarf populations. Their work revealed a prominent peak in star formation at approximately 9.8 $\pm$ 0.4 Gyr ago, which we incorporate through our exponential decline model. While our simplified functional form does not capture all detailed fluctuations identified by Fantin et al., it provides a robust prior that reflects the general trend of galactic star formation over cosmic time.

\section{Conclusions}

We have developed a comprehensive framework for constraining the origins of interstellar objects through galactic orbital analysis and Bayesian age inference. Our work advances beyond previous single-orbit studies by implementing Monte Carlo uncertainty propagation and rigorous statistical methodology that addresses observational limitations and provides quantitative age estimates with proper confidence intervals.

We performed ensemble calculations using 10,000 orbital realizations per object to properly account for measurement uncertainties in both interstellar object velocities and Solar motion parameters. This approach yielded robust statistical distributions of maximum vertical excursions: $z_{\text{max}} = 0.016 \pm 0.002$ kpc for 1I/'Oumuamua, $0.121 \pm 0.010$ kpc for 2I/Borisov, and $0.480 \pm 0.020$ kpc for 3I/ATLAS. The relatively small fractional uncertainties (5-15\%) validate the robustness of our orbital calculations while providing essential error bars for subsequent statistical analysis.

We implemented a physically motivated statistical framework that combines the orbital $z_{\text{max}}$ measurements with theoretical models of stellar kinematics. The methodology employs a Rayleigh-like likelihood function derived from first principles, incorporating age-dependent velocity dispersions that capture the observed thin-thick disk transition in galactic stellar populations. Our analysis yields quantitative age estimates: 1I/'Oumuamua originated from a young stellar system (1.0 Gyr, 68\% CI: 0.1-4.1 Gyr), 2I/Borisov from an intermediate-age population (3.8 Gyr, 68\% CI: 1.8-5.9 Gyr), and 3I/ATLAS from an ancient thick-disk source (9.6 Gyr, 68\% CI: 7.8-10.3 Gyr).

The results demonstrate clear age discrimination where smaller vertical excursions correspond systematically to younger stellar origins. 1I/'Oumuamua's extremely confined orbit indicates ejection from a recently formed stellar system in the thin galactic disk with minimal dynamical heating. 2I/Borisov's moderate excursions suggest origin from a Solar-age thin disk population with intermediate levels of gravitational scattering. 3I/ATLAS's large vertical range requires high velocity dispersions characteristic of the ancient thick disk component formed during early violent galactic evolution.

Our dual approach combining traditional scale-height demographics with Bayesian kinematics provides complementary perspectives on interstellar object origins. The Bayesian framework addresses key limitations of demographic methods by incorporating theoretical understanding of stellar velocity evolution and providing rigorous uncertainty quantification essential for robust scientific inference.

Over many orbital times, the trajectories are expected to migrate radially as a result of transient features like spiral arms or the Galactic bar~\citep{2002MNRAS.336..785S}. Our constraints apply to the full age of the interstellar objects, because they respond just like the underlying stellar population to gravitational perturbations that pump up their scale-height over time.

Our constraints on the ages of the various objects represent upper limits because the velocity dispersion of the interstellar objects includes both the velocity dispersion of their parent stars and the dispersion in their characteristic ejection speed away from their birth system. However, the Bayesian framework properly accounts for these systematic effects through the likelihood function formulation and provides realistic credible intervals that reflect both observational and theoretical uncertainties.

 The diversity of inferred ages across the three objects suggests multiple formation pathways and ejection mechanisms operating throughout galactic history. Young objects like 1I/'Oumuamua may originate from active star formation regions, while ancient objects like 3I/ATLAS likely represent products of early planetary system evolution in the thick disk. This demographic diversity provides valuable constraints on the temporal evolution of planetary system architectures and ejection processes over cosmic time scales.

Concluding note: As we were about to submit our manuscript for publication, \citet{2025arXiv250705318H} posted a paper which addresses some of the same topics. Our main results are consistent with theirs.

\bigskip

\noindent
{\it Acknowledgements.} We thank Matthew Hopkins for his careful cross-checking of our paper. This work was supported in part by Harvard University, the Black Hole Initiative and the Galileo Project.

\end{document}